\documentclass[preprint,showpacs,preprintnumbers,amsmath,amssymb]{revtex4}

\usepackage{graphicx}

\begin{document}

\title{Generating multi-GeV electron bunches using single stage laser wakefield acceleration in a 3D nonlinear  regime}

\author{W. Lu }
\affiliation{Department of  Electrical Engineering, University of California, Los Angeles, CA 90095}
\author{M. Tzoufras }
\affiliation{Department of  Electrical Engineering, University of California, Los Angeles, CA 90095}
\author{F. S. Tsung }
\affiliation{Department of Physics and Astronomy, University of California, Los Angeles, CA 90095}
\author{C. Joshi }
\affiliation{Department of  Electrical Engineering, University of California, Los Angeles, CA 90095}
\author{W. B. Mori }
\affiliation{Department of  Electrical Engineering, University of California, Los Angeles, CA 90095}
\affiliation{Department of  Physics and Astronomy, University of California, Los Angeles, CA 90095}
\author{J. Vieira }
\affiliation{GoLP/Centro de Fisica dos Plasmas, Instituto Superior T\'echnico, 1049-001 Lisboa, Portugal}
\author{R. A. Fonseca }
\affiliation{GoLP/Centro de Fisica dos Plasmas, Instituto Superior T\'echnico, 1049-001 Lisboa, Portugal}
\author{L. O. Silva }
\affiliation{GoLP/Centro de Fisica dos Plasmas, Instituto Superior T\'echnico, 1049-001 Lisboa, Portugal}

\today

\begin{abstract}

 The extraordinary ability of space-charge waves in plasmas to
accelerate charged particles at gradients that are orders of
magnitude greater than in current accelerators has been well
documented. We develop a phenomenological framework for Laser WakeField Acceleration (LWFA) in the 3D nonlinear regime, in which the plasma electrons are expelled by the radiation pressure of a short pulse laser, leading to nearly complete blowout.  Our theory provides a recipe for designing a LWFA for given laser and plasma parameters and estimates the number and the energy of the accelerated electrons whether self-injected or externally injected.  These formulas apply for self-guided as well as externally guided pulses (e.g. by plasma channels).   We demonstrate our results by presenting a sample Particle-In-Cell (PIC) simulation of a $30 fsec$, $200 TW$ laser interacting with a $0.75cm$ long plasma with density $1.5\times10^{18}cm^{-3}$ to produce an ultra-short ($10fs$) mono-energetic bunch of self-injected electrons  at $1.5$ GeV with $0.3 nC$ of charge.   
 For future higher-energy accelerator applications we propose a parameter space, that
is distinct from that described by Gordienko and Pukhov [Physics of Plasmas 12, 043109 (2005)] in that it involves
lower densities and wider spot sizes while keeping the intensity
relatively constant. We find that this helps  increase the output electron beam
energy while keeping the efficiency high.

\end{abstract}

\maketitle

In plasma based acceleration a laser or particle beam creates a plasma wave wakefield with a phase velocity close to the speed of light, c.
\cite{Dawson,Chen}.  The acceleration gradients in these wakefields can easily exceed 10 GeV/m which is nearly three orders of magnitude larger than that achieved in conventional RF technology. A particle injected in such a wave with
sufficient initial energy can interact with the longitudinal
component of the electric field for a long enough time, that its
energy gain is significant.

An intense laser pulse or electron beam can expel the plasma electrons outward to create a bare ion column \cite{sun, RosenzWEIg} which has ideal
accelerating and focusing properties for electron beams
\cite{RosenzWEIg}.
A recent important development in this field has been the simulation
 and experimental observations  of
these highly-nonlinear 3D wakefield structures and the acceleration of electrons
\cite{Pukhov,tsung,Mangles,Geddes,Faure,Hogan}. In the laser-driven cases they were shown to
generate quasi-monoenergetic beams of electrons with energies on
the order of 100 MeV \cite{Mangles,Geddes,Faure}.  In these experiments,
the energetic electron generation mechanism involves an interplay
of complex phenomena, namely: the evolution of the laser, the
self-trapping of electrons and their influence on the wake.

In this article, we
describe how to
 extend the results from the recent experiments towards a more stable regime conducive to making an accelerator. We will show that by using near term lasers  in the $0.1$ to $3 PW$ range
it will be possible to generate of $1-13 GeV$mono-energetic electron beams with ~nC of charge in a single stage without the need for external guiding. Using external guiding energies up to $120 GeV$ can be achieved in a single stage for a $3PW$ laser. Unlike in the recent experiments, this regime is
characterized by lasers whose intensity and spot size are matched and by relatively low plasma densities. The matched laser profile (both transverse and longitudiunal) evolves little during the acceleration distance which  greatly exceeds the Rayleigh length.  


We are motivated to determine a path towards making a compact electron accelerator based on the LWFA concept for use in high energy physics or as a light source. Efficiency and beam quality are key issues one should keep in mind when making such  designs.   As shown by Katsouleas et al.\cite{Katsouleas},  in order to achieve sufficient beam loading efficiency (from wake to particles) and to reduce the beam emittance growth,  the spot size of the wake (hence laser) should be less than a plasma skindepth. In addition,  in order to most efficiently transfer the laser energy to  the wake over the entire effective accelerating distance (pump depletion length matching  dephasing length),  the laser normalized vector potential (laser intensity) must be larger than unity.   These two conditions are based on linear theory, however, taken together they imply that LWFA needs to be operated in a 3D nonlinear regime, in which the laser power exceeds the critical power for relativistic self-focusing\cite{sun}.  

It was shown by Sun et al. \cite{sun} that for intense and narrow lasers the transverse ponderomotive force can lead to complete electron blowout (cavitation)  when the laser power slightly exceeds the critical power for relativistic self-focusing,$P_c$, and the laser is focused tightly. This analysis was done for long pulses (although it assumed the ions were fixed). Mora and Antonsen \cite{Mora} showed in reduced PIC simulation that cavitation could occur for power above $P_c$. When using ultra-short pulses,  $c\tau \lesssim w_{0} \simeq
2\sqrt a_{0}c/\omega_{p}$, where  $a_{0}=
eA_{Laser}/(mc^{2})$ for a linearly polarized laser and $w_0 $ the laser spot size, the ions are indeed stationary, and
a region nearly void of electrons (ion channel) is generated behind the laser
pulse. This channel exerts an attractive Coulomb force on the
blown-out electrons causing them to rush back toward the axis thereby exciting a wakefield. In 1991 Rosenzweig et al. \cite{RosenzWEIg} showed that wakes excited in  this nonlinear blowout regime (they used a particle beam driver) had ideal focusing and accelerating properties. Very soon thereafter it was shown using 2D simulations that similar wakes could be made by lasers \cite{mori_blowout}. 

 In 2002 Puhkov and Meyer-ter-vehn used 3D particle-in-cell simulations of LWFA at very high laser intensities and plasma densities without external guiding and found that electrons were self-injected in the blowout regime and that they formed a quasi mono-energetic beam as they were accelerated \cite{Pukhov}.  In their simulations the blowout cavity formed a sphere leading them to subsequently call this the ``bubble" regime for LWFA \cite{kostyukov}.  Recently, a theoretical analysis  \cite {WEI,Wlpop} showed that for sufficiently intense and ultra-short lasers, $a_{0} \gtrsim  4$,  the resulting electron density
structure will resemble a spherical cavity and it was shown that the laser spot size needed to be matched to the maximum blowout (bubble)  radius in order to achieve the most ideal wakes. It is worth noting for $ 2\lesssim a_0\lesssim 4$ electron blowout still occurs with the cavity slightly deviating from a spherical shape.

The recent experimental results are promising but they still rely on substantial evolution of the laser and the energy spread and emittances are not sufficient for high-energy physics applications. What paths are available for extending these experiments towards higher energies with more control using the ``bubble" regime? The original simulation results on the ``bubble" regime in ref. \cite{Pukhov} were based on a  simulation of a $360 TW$ laser propagating through a $0.01$ critical density ($1.74\times10^{19} cm^{ -3}$ for $\lambda=0.8\mu m$) plasma. Do the key features of this simulation remain for different laser powers, pulse lengths, spot sizes,  and plasma densities? Answering these questions is difficult because the inherent physics is highly nonlinear. 

One approach was put forth by Gordienko and Puhkov (GP) \cite{Gordienko}. They argued that for $a_0>>1$ the speed of all plasma electrons is very close to the speed of light and under this condition all quantities will scale with a single similarity parameter, $S\equiv{n_p\over {n_c a_0}}$, to some power. The coefficients in front of the scalings are determined from simulations (or experiments). The resulting expressions are therefore only strictly valid so long as the laser's transverse and longitudinal profile, aspect ratio etc. remain the same. Another approach presented here is to use a phenomenological description. We identify the important physics as wake excitation (amplitude and phase velocity), pump depletion (pulse evolution), dephasing (between particles and wake),  and beam loading. We use these concepts to develop expressions for predicting the number of electrons, the electron energy, and overall efficiency.

For any useful accelerator, the accelerating structure's (wake) amplitude, wavelength, and phase velocity need to be reasonably stable during the entire acceleration process.  We show that a stable wake can be excited  in the blowout regime  when the  laser's spot size is roughly matched to the blowout radius. We also show that under these conditions expressions for the wake amplitude and phase velocity (dephasing length), the laser pump depletion length, and the number of accelerated electrons can then be obtained including their coefficients based on simple physics arguments and previously published ideas. Importantly, unlike when using a similarity theory,  the coefficients derived from the phenomenological theory can be extended to propagation in plasma channels as well to when an external beam of electrons is injected. 

We will show that while there is overlap in the ideas put forth here and by GP there are also profound differences besides those just described. Their analysis can only be  valid when $a_0 \gtrsim2 \sqrt{n_c/n_p}$ while ours is for $a_0\gtrsim2$.  Their supporting simulations used $a_0=20-80$ (for each direction of a  circularly polarized laser which is the same as  $28-113$ for a linearly polarized laser) and $n/n_c=0.02-0.08$. Their analysis appears to ignore dephasing leading  to different scaling laws. We will show that in contrast to the work of GP, our expressions advocate using lower $a_0$ and lower densities. In fact, we will argue that in order to obtain controlled acceleration one should operate with an $a_0$ too low for the key assumption in the  theory of GP to even apply. Despite these differences we borrow the term ``bubble" from ref. \cite{kostyukov} to describe the near spherical ion channel behind the laser pulse. 

To verify our results, we have carried out many computer experiments using the 3D,
particle-in-cell code OSIRIS \cite{Fonseca}, that explore a wide
range of plasma densities, laser powers, and spot sizes (see
later). We highlight one simulation that is very relevant to
near term experiments. Details of the simulation parameters and setup are provided in the Appendix. In this simulation, a $30 fs$ (FWHM) $0.8
\mu m$ laser pulse containing $200$ TW of power is focused to a
spot size $w_{0} = 19.5 \mu m$ at the entrance of a $1.5 \times
10^{18} cm^{-3}$ density plasma to give a normalized vector
potential of $a_{0} = 4$. The laser is circularly polarized (with normalized vector potential $4/\sqrt{2}$ in each direction) and
has a Gaussian transverse profile. The plasma is $0.75cm$ long which corresponds to more than $5$ Rayleigh lengths.

We next develop a phenomenological theory for LWFA in the blowout regime. A light pulse can be guided with a nearly constant (matched) spot size by a plasma channel that has a
parabolic refractive index/density profile with a maximum/minimum
on axis.  The index of refraction in  a plasma can be expanded as\cite{Esarey, Mori}  $ \eta = ck/\omega \simeq 1-\frac{1}{2}\frac{\omega_{p}^2}{\omega_{0}^{2}}\biggl(1+\frac{\Delta n_c}{n_p} \frac{r^2}{w_0^2} + \frac{\Delta n}{n_p} -\frac{a_0^2}{8}\biggr)$  where $\Delta n_c$  parametrizes an external density channel, $\Delta n$  is a density depletion from the transverse ponderomotive  force and the term $a_0^2/8$ is due to relativistic mass corrections.  The characteristic density change required to optically guide such a profile with little spot size oscillation is $\Delta n_{c}=1/(\pi{r}_{e}w_{0}^{2})$ \cite{Esarey},
where $r_{e}=e^{2}/mc^{2}$ is the classical electron radius and
$w_{0}$ is the laser spot size. If the density depression is normalized to the plasma density $n_p$, this condition becomes, $\Delta n_{c} / n_{p} \simeq 4/(k_{p}w_{0})^{2}$. 
An equivalent change to the index of refraction from relativistic mass corrections can also self-guide
a laser if $a_0^2/8\gtrsim4/(k_p w_0)^2$ or $P\gtrsim P_c$, where $P_{c} = 17 \omega_{0}^{2}/\omega_{p}^{2}[GW]$ is the
critical power for relativistic self focusing \cite{sun}.

 Unfortunately, even when $P\gtrsim P_c$ and the spotsize is sufficiently small,  the index of refraction  needs a distance on the order of $c/\omega_p$ to build up. 
 Therefore, it is often thought that a short pulse laser $\tau\lesssim 1/\omega_p$ cannot be self-guided and some form of 
external optical guiding is needed \cite{Esarey}. However, as described in ref.
\cite{deckermori} for $ P/P_{c}>>1$ a degree of self-guiding for
short pulses is possible because the leading edge of the laser
locally pump depletes before it diffracts and the back of the pulse is still guided in the ion column region. In case of external optical guiding $ P/P_{c}>>1$ is not necessary, instead $P/P_c\gtrsim1$ can be used which also leads to clear blowout.

\begin{figure}[!tb]
\begin{center}
\includegraphics[width=3.4in]{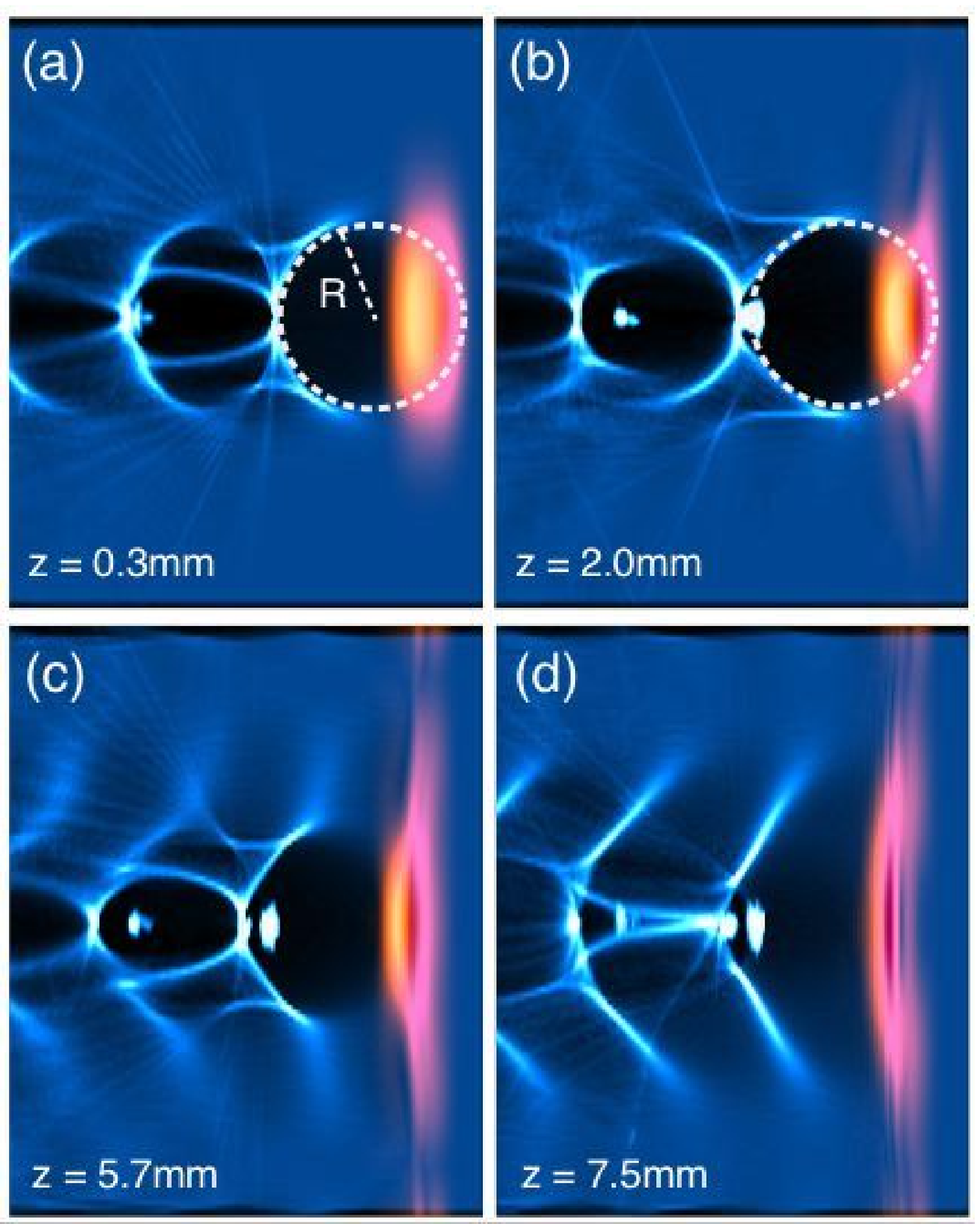}
\caption{A sequence of 2-dimensional slices $(x-z)$ reveals the evolution of the accelerating structure (electron density, blue) and the laser pulse (orange). Each plot is a rectangular of size $z = 101.7\mu m$ (longitudinal direction, $z$) and $x=129.3\mu m$ (transverse direction, $x$).  A broken white circle is superimposed on each plot to show the shape of the blown-out region. When the front of the laser has propagated a distance, (a) $z = 0.3mm$  the matched laser pulse has clearly excited a wakefield. Apart from some local modification due to beam loading effects, as seen in (b)
  this wakefield remains robust even as the laser beam propagates though the plasma a distance of $7.5mm$  (as seen in (c) and (d)) or 5 Rayleigh lenghts. After the laser beam has propagated $2mm$ (as seen in (b)) into the plasma, one can clearly see  self-trapped electrons in the first accelerating bucket. The radial and longitudinal localization of the self-trapped bunch is evident in part (c). After $7.5mm$  the acceleration process terminates as the depleted laser pulse starts diffracting.} \label{figure1}
\end{center}
\end{figure}

The stable self-guiding of an intense  short pulse  is illustrated in
figure \ref{figure1}. We find that for  self-guided propagation of the laser beam,
without significant variations of the pulse profile over the
interaction distance, its spot size and intensity must be
appropriately chosen.
A rigorous derivation  for the  matched spot size and the laser profile is not currently
available. Still, as noted in ref.\cite{sun, kostyukov,WEI} the
requirements for a matched profile can be estimated by assuming
that the transverse ponderomotive force of the laser $k_{p} \nabla
a_{0}^{2}/\gamma\sim a_{0}/(k_{p}R)$, where $R$ is the blowout radius,   is roughly balanced by the
force of the ion channel $E_{r}\sim k_{p}R$, which pulls back the
ponderomotively expelled electrons.  Equating these two
expressions yields $k_{p}R\sim \sqrt{a_{0}}$. Furthermore, the blowout radius scales with the laser spot size,  $k_{p}w_{0} \sim k_{p}R$. Through
simulations we have found that a more refined condition which leads to only slight oscillations in the spot size is:
\begin{equation}\label{matching}
k_{p}R \simeq k_{p} w_{0} =2\sqrt{a_{0}}
\end{equation}
We find that this relationship still holds for $a_{0}\gtrsim 2$. 
In the work of Gordienko and Pukhov \cite{Gordienko} it was assumed $k_{p}R \simeq k_{p} w_{0} =1.12\sqrt{a_{0}} $ (we rewrote this for a linearly polarized laser).
We can use Eq. \ref{matching}  to reformulate  the matched beam spot size
condition as:   
\begin{equation}\label{matching2}
a_{0} \simeq 2 (P/P_{c})^{1/3}
\end{equation}

As alluded to earlier, the laser etches back due to local pump
depletion. In ref.\cite{deckermori}, an estimate of the etching
rate, i.e., the etching velocity, based on nonlinear 1D effects was given as
$\upsilon_{etch} \simeq c \omega_{p}^{2}/\omega_{0}^{2}$; and it
was verified in 2D PIC simulations. We find that on average this estimate
also agrees very well with the observations in numerous 3D PIC simulations.
Therefore, the laser will be depleted after a distance (pump depletion length)
\begin{equation}\label{pump_depletion} L_{etch} \simeq
\frac{c}{\upsilon_{etch}} c\tau_{FWHM}\simeq
\frac{\omega_{p}^{2}}{\omega_{0}^{2}}c\tau_{FWHM}
\end{equation}
The front of the laser,
that excites the wake, moves backward as the pulse etches back with
$\upsilon_{etch}$. The phase velocity of the wake can therefore be
expressed as $\upsilon_{\phi} \simeq \upsilon_{g} -
\upsilon_{etch}$, where $\upsilon_{g}$  is the linear group
velocity of light in a very underdense plasma $\omega_{p}^{2} \ll
\omega_{0}^{2}$; therefore $\upsilon_{\phi} \simeq
c[1-3\omega_{p}^{2}/(2\omega_{0}^{2})]$. The distance that the
trapped electrons travel until they outrun the wave (dephasing length) is:

\begin{equation}\label{dephasing}
L_{d} \simeq \frac{c}{c-\upsilon_{\phi}} R  \simeq
\frac{2}{3}\frac{\omega_{0}^{2}}{\omega_{p}^{2}}R
\end{equation}

We find in numerous 1D, 2D and 3D simulations  that the etching velocity and hence this dephasing estimate works well for $2\lesssim a_{0} \lesssim 2 \sqrt{n_c\over n_0}$. The estimate for the upper value of $a_0$ is discussed later.

For the energy gain we may write the obvious equation
\begin{equation}\label{energygain}
\Delta E = q E_{LW} L_{acc} = \epsilon_{LW} l_{acc} mc^2
\end{equation}
where $E_{LW}$ is the average accelerating field of the beam
loaded wake, $L_{acc}$ is the acceleration length,
$\epsilon_{LW}\equiv eE_{LW}/(mc\omega_{p})$ and
$l_{acc}=\omega_{p} L_{acc}/c$.
The desired acceleration length is the dephasing length, so we
impose the condition $L_{etch}>L_{d} \Rightarrow c\tau_{FWHM} >
2R/3$. If the pulse is too short dephasing will not be reached,
and the electron beam may have significant energy spread.
We note that this condition is approximate, since the laser
starts diffracting as soon as its intensity is insufficient to
sustain self focusing and this happens before the pulse is
completely pump depleted. Additionally the injected particles need
to slightly pass the dephasing point (phase space rotation) so
that the energy spread is minimum. On the other hand, the length
of the pulse should not be too large,  because the laser field
could interact with the trapped electrons and degrade the beam
quality.

\begin{figure}[!tb]
\begin{center}
\includegraphics[width=3.4in]{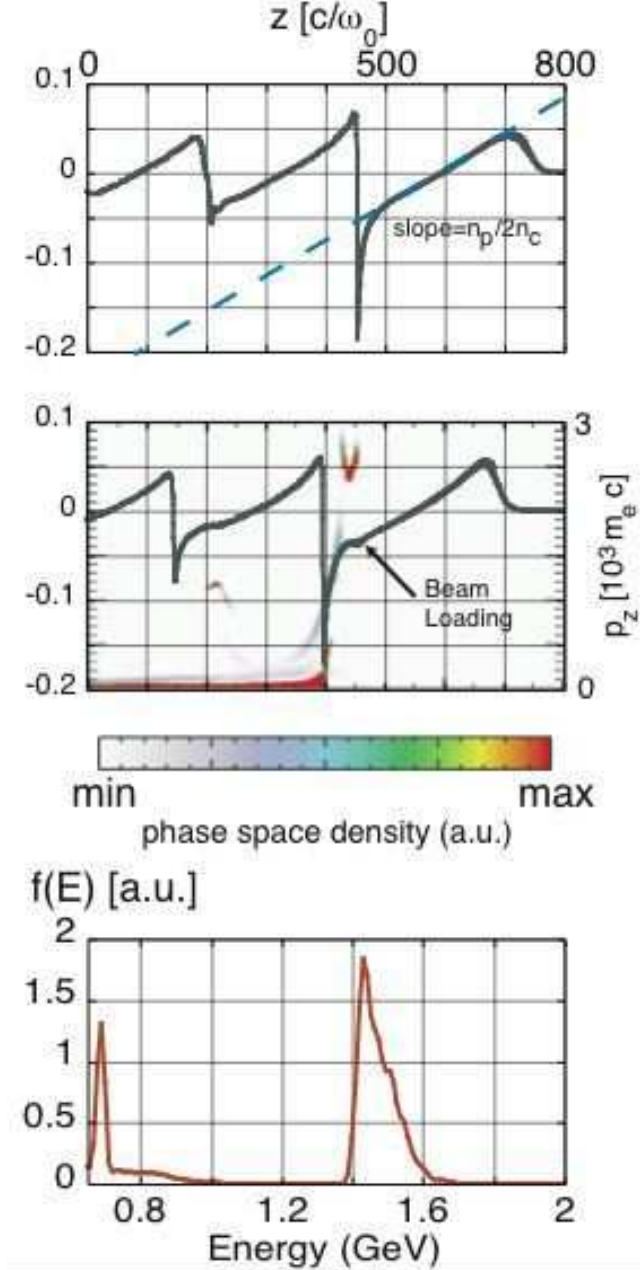}
\caption{(a) A lineout of the wakefield along the $z$ axis after $0.3 mm$ shows that within the first bucket the slope of the wakefield is nearly constant and equal to $eE_{z}/(mc\omega_{p}) \simeq \xi/2$ where $\xi = (k_{p}(ct-z))$. After $5.7 mm$ of propagation (b) the wakefield has been modified by beam loading (flattening of the wake between $400-450c/\omega_{p}$). This is corroborated by the $p_{z}$ vx $z$ plot that  is superimposed on the lineout of the wakefield. Pictures (a) and (b) reveal that the acceleration mechanism is extremely stable during the simulation. The energy spectrum after $7.5mm$ (c) exhibits an isolated spike of $0.3nC$ at $1.5GeV$ with energy spread $\Delta \gamma/\gamma =3.8\%$ corresponding to the first bucket and a second spike of $50pC$ at  $700MeV$ with energy spread  $\Delta \gamma/\gamma = 1.5\%$ corresponding to the second bucket.} \label{figure2}
\end{center}
\end{figure}

We need an expression for $\epsilon_{LW}$ which is the
average accelerating field experienced by an electron, $\langle E_{z}\rangle$.
In ref. \cite{WEI} it was shown that for $a_{0} \gtrsim 4$  the ion column
formed a sphere and that the accelerating field, for the most part, depends linearly on the distance from the middle of the sphere (figure \ref{figure2}(a)). This is confirmed in figures 1(a)-(c) where a bubble void of electrons
roughly forms a circle and in figures 2(a),(b) where a lineout of
$eE_{z}/(mc\omega_{p})$ along the axis is shown.
We also found by theory and 3D PIC simulations
that a spherical shape is still roughly formed for $2\lesssim a_0 \lesssim 4$ with matched laser profiles.  
 Interestingly, although the
physics is very different, the nonlinear wakes formed in 1D also
exhibit a linear slope of the same amount.  

Because the bubble is roughly a sphere and the electrons are either self or externally injected at the rear,  the electrons then travel a relative distance  R before they dephase. The peak useful accelerating field is $eE_{z,max}/(mc\omega_p)=  \sqrt{a_0}$ and because the wakefield is roughly linear,  the average field is half of the peak: $\epsilon_{LW}
\equiv e{E}_{z,max}/(2mc\omega_{p})\simeq \sqrt{a_{0}}/2$ . We can
therefore write the approximate equation for the energy gain:

\begin{eqnarray}\label{energy_gain}
  \Delta E  &\simeq&\frac{2}{3}mc^{2}\biggl(\frac{\omega_{0}}{\omega_{p}}\biggr)^{2}a_{0} \nonumber\\
  &\simeq& mc^{2}\biggl(\frac{P}{m^2c^5/e^2}\biggr)^{1/3}\biggl(\frac{n_{c}}{n_{p}}\biggr)^{2/3} \nonumber\\
 \Delta E  [GeV] &\simeq& 1.7\biggl(\frac{P[TW]}{100}\biggr)^{1/3}\biggl(\frac{10^{18}}{n_{p}[cm^{-3}]}\biggr)^{2/3}
  \biggl(\frac{0.8}{\lambda_{0}[\mu m]}\biggr)^{4/3}
 \end{eqnarray}

We  emphasize the much stronger dependence of the beam
energy on the plasma density than on the input laser power.
However, when the plasma density is lowered for fixed power, ensuring self-guided propagation of the leading edge of the laser is more challenging. This can be accomplished by  plasma channels or to some degree by self-guiding. As we argue later, for self-guiding to occur
 $P/P_c$  needs to increase as the plasma density
decreases. 
We can rewrite equation (\ref{energy_gain}) in terms of the
critical power for relativistic self-focusing, $P_{c}$:
\begin{equation}\label{energy_power}
 \Delta E  [GeV] \simeq 3.8
\biggl(\frac{P}{P_{c}}\biggr)^{-2/3}\frac{P[TW]}{100}
\end{equation}
 On the other hand, if a plasma channel is used, $P/P_c$ can be kept as low as $1$. As shown before,  the channel depth $\Delta{n}_c/n_p$ needed is $4/(k_p w_0)^2\simeq1/a_0$. So as long as $a_0\gtrsim2$ is used, the normalized channel depth is small (less than $0.5$). It is also worth noting here that when a channel is used, the channel parameters (width and depth) should be chosen based on the matching condition for given laser power and plasma density (Eqs. \ref{matching} and \ref{matching2}) so that leading front of the laser is guided by the density channel and the back of the laser is guided by the matched ion channel.

The electron beam surfing on the wake can be either self-injected as shown in our  sample simulation  (figure \ref{figure1}(b)) or externally injected from some other source. For self-injection, particles in
the rear of the blowout region must  be able to catch up with the
wake. The physical condition for this  to happen is  twofold: first, the blowout radius should be large enough so that when the particles reach the rear of the bubble, they move predominately in the forward direction with speed close to the speed of light. Second, at the rear portion of the ion channel, trajectory crossing occurs leading to a narrow sheath with the highest accelerating and focusing fields. Therefore, even though electrons initially have a $\gamma$ (energy) substantially below the
wake's Lorentz factor $\gamma_\phi$, they can  easily achieve sufficient energy as they are accelerated
while they slowly drift backwards (relatively to the pulse) in the sheath.  In our sample simulation, the effective $\gamma$  of the wake is around $20$ and the normalized blowout radius is around $4$.  The initial energy $\gamma$ of those trapped electrons is substantially smaller than $20$. For even lower plasma densites, we have performed a number of simulations, where
an electron beam with $\gamma$ exceeding 10000 was used as the
driver instead of a laser  and we observed self-injected electrons
in each case for a normalized blowout radius around $5$. This indicates that for  laser wavelengths in the $0.8 \mu m$ range and plasma densities of interest, self-injection will always happen when we keep the normalized  blowout radius around $4\sim5$. This differs significantly from the claim in ref. \cite{kostyukov} that $\sqrt{a_0}>\gamma_\phi$, or $a_0>400$ for our sample simulation, for self-trapping to occur.

\begin{figure}[!b]
\begin{center}
\includegraphics[width=3.4in]{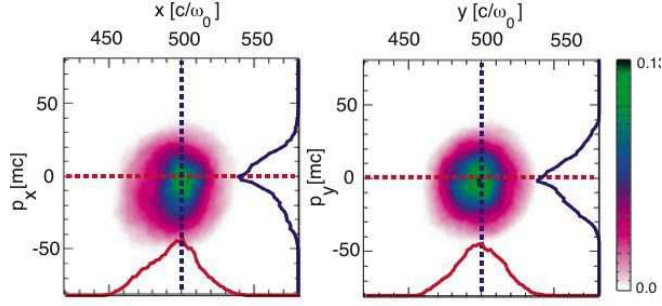}
\caption{ The normalized
emittance $[\varepsilon_{N}]_{i} =\pi \sqrt{\langle\Delta p_{i}^{2}\rangle \langle\Delta x_{i}^{2}\rangle-\langle\Delta p_{i}\Delta x_{i}\rangle^{2} }$ (where $\Delta p_{i}$ is normalized as indicated by the figure and the emittance is in units of $\Delta x_{i}$) and is the approximate area in phase space $p_{i}x_{i}$.  For the pictures above which correspond to the first bunch this formula yields: $[\varepsilon_{N}]_{x}\simeq 35  \pi \cdot mm \cdot mrad$ and  $[\varepsilon_{N}]_{y}\simeq 29 \pi \cdot mm  \cdot mrad$. An upper limit for the emittance can be found by multiplying the typical divergences shown in the figure; this method leads to an overestimation which for this case is about $25\%$. For the second bunch of accelerated electrons (not shown in this figure) the emittances are significantly lower: $[\varepsilon_{N}]_{x}\simeq 10  \pi \cdot mm \cdot mrad$ and  $[\varepsilon_{N}]_{y}\simeq 11 \pi \cdot mm  \cdot mrad$.
} \label{figure4}
\end{center}
\end{figure}

In the regime presented here the self-injected electron  bunches are
highly localized in space with a half-width of the first bunch of
only $\sim10 fs$, i.e. $1 c/\omega_{p}$. Once a sufficient number of electrons have been
trapped the trapping process terminates, as seen in figure \ref{figure1}(c).  The
first electron bunch reaches an energy of 1.5 GeV and its energy
spectrum is presented in figure \ref{figure2}(c). The normalized emittances are shown in figure \ref{figure4}.  They may be estimated as  the product of the beam spot size, which roughly scales with $1/\sqrt{n_p}$, with the spread in the momentum perpendicular to the acceleration direction, which scales with the relativistic ponderomotive potential ($\sim a_0$).  These simple considerations show that as we move to lower densities in order to achieve higher energy particles the emmitances of the self-injected electrons will increase. This suggests that for the electron beam to be useful for high energy physics or light source,  external injection may be more attractive. As an interesting aside, simulations also reveal 
the trapping and acceleration of a second distinct bunch in the
second bucket (see figure \ref{figure2}(c)), which has a lower energy because the average
electric field it experiences is smaller.

The number, $N$, of electrons that are accelerated can be estimated
from energy balance. Hence we examine the partition of field and particle energy within the first bucket. The fields inside the ion column have $E_{z}$, $E_{r}$, and $B_{\phi}$ components. In addition there is kinetic energy in the plasma. In a 3D linear or 1D nonlinear wake the fields and kinetic energy scale together. This is not the case for these 3D nonlinear wakes where an increasing percentage of energy ends up in electrons which are blown out well beyond the narrow electron sheath for higher  laser intensity.  Integrating the field energy in the ion channel we find equipartition between the energy in the longitudinal field $\mathcal{E}_{l}$ and the focusing fields $\mathcal{E}_{f}$: 
\begin{equation}\label{energy_in_fields}
\mathcal{E}_{l}\simeq\mathcal{E}_{f}\simeq
\frac{1}{2}\mathcal{E} = \frac{1}{120}(k_{p}R)^{5}\biggl(\frac{m^{2}c^{5}}{e^{2}\omega_{p}}\biggr)
\end{equation}

The trailing particles can recover the field energy in the ion channel and the kinetic energy in the narrow electron sheath by changing the shape of the ion channel. The kinetic energy in the narrow electron sheath scales the same way as the field energy.  By equating $2\mathcal{E}$ with the energy absorbed by $N$ particles that travel across the ion channel (we assume the average field felt by these particle is $E_{z,max}/2$ and the kinetic energy in the electron sheath absorbed by the trailing particles is the same as the field energy), we obtain:
\begin{equation}\label{number_particles}
N\simeq \frac{1}{30}(k_{p}R)^{3}\frac{1}{k_{p}r_{e}}
= \biggl(\frac{\beta^{3}}{\alpha}\biggr)\frac{8/15}{k_{0}r_{e}}\sqrt{\frac{P}{m^2c^5/e^2}}
\end{equation}
where  $\alpha = k_{p}w_{0}/(2\sqrt{a_{0}})$ and $\beta = k_{p}R/(2\sqrt{a_{0}})$. Using equation (\ref{matching}) $\alpha\simeq 1\simeq \beta$ we obtain:
\begin{equation}\label{particle_scaling}
N\simeq \frac{8/15}{k_{0}r_{e}}\sqrt{\frac{P}{m^2c^5/e^2}}\simeq 2.5\cdot 10^{9}\frac{\lambda_{0}[\mu m]}{0.8}\sqrt{\frac{P[TW]}{100}}
\end{equation}

The efficiency scales as the total energy $\mathcal{E}_{b}$ in the
accelerated electron beam (energy gain equation (\ref{energy_gain}) times
particle number from equation (\ref{number_particles})) devided by the total
laser energy $\mathcal{E}_{T}$ (assuming $c\tau \simeq 2\sqrt
a_{0}c/\omega_{p}$):
\begin{equation}\label{efficency}
\Gamma \sim \mathcal{E}_{b}/ \mathcal{E}_{T} \sim 1/ a_{0}
\end{equation}
which indicates that $a_{0}$, i.e.,  $(P/P_{c})^{1/3}$ cannot be too large if one needs
high efficiency. For a $200TW$, $0.8 \mu m$ pulse equation (\ref{particle_scaling}) predicts $0.6nC$ of charge. The charge measured from the simulation for the first bunch is $0.3 nC$. We have also verified these scaling laws by monitoring how many electrons can be externally injected before the wake becomes severely loaded. 

As shown in figure \ref{figure1}(d), the acceleration
process stops before the accelerating bunch dephases.
This will not lead to any considerable modifications of the
aforementioned formulas, particularly because the pump depletion
length scales as the dephasing distance and  the
accelerating wakefield decreases as the trapped electrons approach
the center of the sphere.

The beam energy seen in the simulation, 1.5 GeV,  is close to that calculated
theoretically  from Eq. (\ref{energy_gain}) which is
$(\ref{energy_gain}) \Rightarrow \Delta W \simeq 1.6 GeV$. Using
formulas  (\ref{pump_depletion})-(\ref{dephasing}) we see
$L_{d} \simeq 1.31cm > 0.96cm\simeq L_{pd}$ which is in
agreement with our observation that pump depletion happened before
dephasing. In addition, as the laser pump depleted its rate of diffraction increased such that the effective acceleration length was eventually limited by diffraction to a distance
 $0.75cm$ which was less than the pump depletion distance $0.96cm$. For a wake with a linear slope most of the energy gain occurs near the peak field. This combined with the density spike at the back of the bucket is why the estimate from  eq. (\ref{energy_gain}) is still so accurate for this non-optimized sample simulation.

\begin{figure}[!tb]
\begin{center}
\includegraphics[width=3.4in]{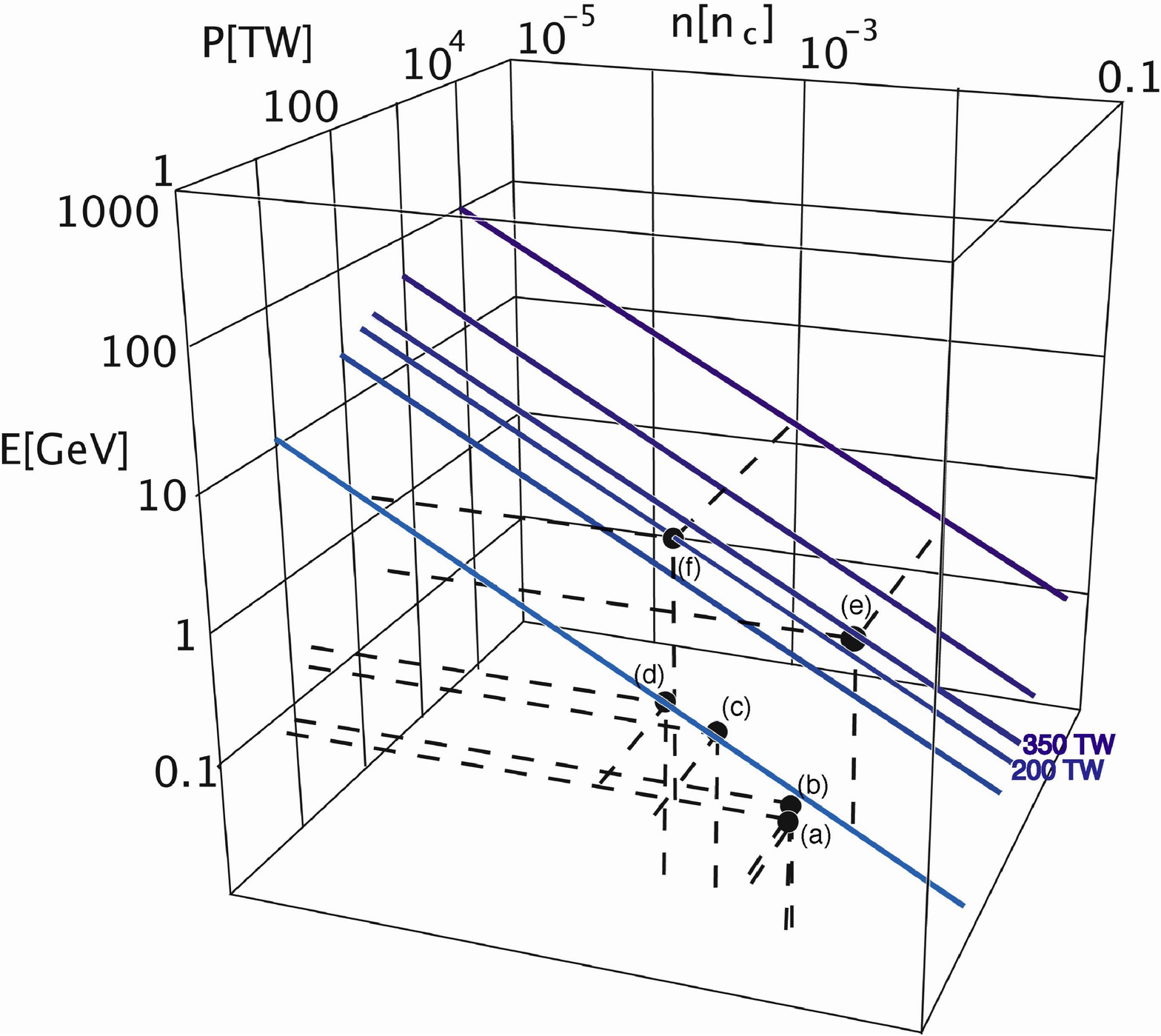}
\caption{$E[GeV]$ vs. power $P[TW]$ and density $n/n_{c}$ from equation  \ref{energy_gain} : The blue lines of constant power show the strong dependence of the energy of the self-trapped electrons to the density. The black points correspond to:
(a) experiment \cite{Mangles}, (b) experiment \cite{Geddes} which uses a channel for guiding, (c) experiment and 3D PIC simulation \cite{Faure}, (d) 3D PIC simulation \cite{tsung} which uses a channel for guiding, (e) A 3D PIC simulation in \cite{Pukhov}, (f) 3D PIC simulation presented in this article. Each of these points is very close to one of the blue lines indicating agreement with our scaling law.} \label{figure3}
\end{center}
\end{figure}

In spite of the complexity of the physics associated with this
interaction, the predictions by the simple formulas presented in
this article are very close to 3D PIC simulation results. Good agreement is also achieved between these scaling laws and recent experimental  results \cite{Mangles,Geddes,Faure} despite the fact that the laser powers were slightly below the ``threshold"  for  the blowout regime and the laser pulse was not matched transversely nor longitudinally. To be in the regime identified by a spherical bubble, one needs  $a_{0} \gtrsim 4$ or equivalently $P/P_{c} \gtrsim 8$ and $c\tau < 2 \sqrt{a_0} c/ \omega_{p}$ which leads to the condition that $P \gtrsim 30 (\tau/{30fs}) TW$. These conditions can be relaxed for channel-guided lasers for which $P/P_{c}$ may be smaller. It is then written as     $a_{0} \gtrsim 2$ or  $P/P_{c} \gtrsim 1$. We present the comparison between the scaling law for the energy Eq. (\ref{energy_gain}) and the aforementioned results in figure \ref{figure3}.

The scalings derived above and the
underlying physics predict that it is advantageous to use moderate
intensities and very low plasma densities to
increase the output energy and keep the efficiency high.  The simulations also show that the
injection process can be clamped and the energy spread of the
electron beam is much less in the regime we are proposing, which indicates that this
regime is also amenable to accelerating externally
injected beams while maintaining good beam quality.

\begin{table}[!tb]
\begin{center}
\begin{tabular}{r|c|c|c|c|c|c|c|c|}
\cline{2-9}
&$a_0$&$k_p w_0$&$\epsilon_{LW}$&$k_pL_d$&$k_pL_{pd}$&$\lambda_W$&$\gamma_\phi$&$\Delta W/(mc^2)$\\
\cline{2-9}
Linear:  &$<1$&$2\pi$ &$a_0^2$&$\frac{\omega_0^2}{\omega_p^2}$&$\frac{\omega_0^2}{\omega_p^2}\frac{\omega_p \tau}{a_0^2}$&$\frac{2\pi}{k_p} $&$\frac{\omega_0}{\omega_p}$&$a_0^2\frac{\omega_0^2}{\omega_p^2}$\\
\cline{2-9}
1D Nonlinear:&$>1$&$2\pi $&$a_0$&$4a_0^2\frac{\omega_0^2}{\omega_p^2}$&$\frac{1}{3}\frac{\omega_0^2}{\omega_p^2}\omega_p \tau$&$\frac{4a_0}{k_p} $&$\sqrt{a_0}\frac{\omega_0}{\omega_p}$&$4 a_0^2\frac{\omega_0^2}{\omega_p^2}$\\
\cline{2-9}
3D Nonlinear:&$>2$&$2\sqrt{a_0} $&$\frac{1}{2}\sqrt{a_0}$&$\frac{4}{3}\frac{\omega_0^2}{\omega_p^2}\sqrt{a_0}$&$\frac{\omega_0^2}{\omega_p^2}\omega_p \tau$&$\sqrt{a_0}\frac{2\pi}{k_p}$&$\frac{1}{\sqrt{3}}\frac{\omega_0}{\omega_p}$&$\frac{2}{3}\frac{\omega_0^2}{\omega_p^2}a_0$\\
\cline{2-9}
ref. \cite{Gordienko}: &$>20$&$\sqrt{a_0} $&$\sqrt{a_0}$&$\mbox{ }$&$a_0\frac{\omega_0^2}{\omega_p^2}\omega_p \tau$&$\mbox{ }$&$\mbox{ }$&$\frac{\omega_0^2}{\omega_p^2}a_0^{3/2}\omega_p\tau$\\
\cline{2-9}
\end{tabular}
\caption{This table compares formulas and scalings from linear theory, 1D nonlinear theory, the similarity theory of reference 16 and the 3D phenomenological nonlinear theory in this paper. The column labeled $a_0$ shows the range of $a_0$ that each theory is intended for. $k_pw_0$ from linear and 1d nonlinear theory is kept constant while in both our theory and theory of ref. \cite{Gordienko}. $k_pw_0$ scales with the blowout radius. The acceleration length is limited by dephasing in linear theory and our 3D nonlinear theory, while it is limited by pump depletion in 1D nonlinear theory and the work of GP. (Note that in standard 1D nonlinear theory the phase velocity of the wake is incorrectly set equal to the nonlinear group velocity) Only in our 3D nonlinear theory do dephasing and pump depletion scale the same way (if the pulse length is matched to the bubble size).  The dephasing length is evidently ignored in ref. \cite{Gordienko} so the energy gain depends on the pulse length. Linear dephasing and 1d nonlinear dephasing both scale differently than in the 3d nonlinear regime. The pump depletion length from 1d nonlinear theory scales as in 3d nonlinear theory but for different physics reasons. In 1D all the laser energy goes into the wake while in our 3D theory this is not the case. We discuss the pump depletion length from ref. \cite{Gordienko} extensively in the text. Both linear and 1d nonlinear theory incorrectly identify the plasma wavelength and wake phase velocity. The resulting energy scalings are different. If one sets $\omega_p\tau\sim\sqrt{a_0}$ in ref. \cite{Gordienko} then the energy scales as $a_0^2$. Therefore linear, 1D nonlinear and the similarity theory of ref. \cite{Gordienko} all give identical scalings for the $\Delta W$ with both intensity and density while our results give $\Delta$ scaling as $a_0$ to the first power.}\label{table1}
\end{center}
\end{table}

In table \ref{table1} we compare some of our results to previously published work \cite{Gordienko,Esarey}. Both linear and 1D nonlinear formulas fail to describe the 3D nonlinear regime in nearly all respects. It is however instructive to compare the scaling laws described above against those obtained in ref.\cite{Gordienko}
using a similarity theory for the high intensity high plasma density regime, because in ref. \cite{Gordienko} the scaling (but not the coefficients) of the fields in the bubble and the laser spotsize with $a_0$ were identified correctly. The key assumption
of ref.\cite{Gordienko} required  that $a_{0}>>1$. For the energy gain,
they obtained
\begin{equation}\label{similarity}
E_{mono}\approx0.65mc^2\sqrt{\frac{P}{m^2c^5/e^2}}\frac{c\tau}{\lambda}
\end{equation}
This formula also implies a dependence on plasma density if 
 $c\tau\approx{w}_{0}\approx\sqrt{a_{0}}c/\omega_{p}$ is assumed. Under these optimum conditions, formula (\ref{similarity}) can be rewritten as
\begin{equation}\label{similarity1}
 E_{mono}\approx 0.16 mc^2\frac{c\tau}{w_{0}}\biggl(\frac{P}{m^2c^5/e^2}\biggr)^{2/3}\biggl(\frac{n_{c}}{n_{p}}\biggr)^{1/3}
\end{equation}

 We point out the stronger dependence of the energy on the power in these formulas.
The difference between the scaling laws presented here and those from similarity theory lie in the scaling for $L_{acc}$.   We have argued that $L_{acc}$ is limited because the pulse pump depletes primarily by giving kinetic energy to the electrons. An electron at the front of the laser is pushed forward and to the side. As argued in ref. \cite{deckermori} each electron gains an amount of energy that scales as $(a_{0}^{2}/2)mc^2$. Since the laser's energy also scales as $a_{0}^2$ this ``1D like" pump depletion 
enhances the etching velocity independently of laser intensity. However, for extremely large $a_{0}$ electrons can move forward with a velocity greater than the velocity of the leading edge of the pulse; which is at most the linear group velocity because the nonlinearity does not develop instantaneously. Equating the forward going velocity of a single electron to the linear group velocity gives an estimate for a critical value for the laser amplitude,  $a_{0c}\approx2\sqrt{n_{c}/n_{p}}$. We can see that for high
plasma density  $n_{p}/n_{c}\approx0.01\sim0.08$ the critical value for $a_{0}$ is 
$a_{0c}\approx20\sim8$, but for lower plasma densities 
$n_{p}/n_{c}\sim0.001$, $a_{0c}\gtrsim70$. 

For laser amplitudes above this critical value, the laser will pump deplete more slowly  because the kinetic energy given to the each electron will no longer scale as $a_{0}^2$. In this limit a constant percentage of the energy may go into the fields of the wake such that  $\Gamma\sim\mbox{constant}$ as predicted by ref. \cite{Gordienko}. The pump depletion length, $L_{pd}$, can then be estimated from energy balance between the laser and the wakefields, $\frac {E_{0}^2}{8\pi} c\tau \approx \frac {E_{z}^2}{8\pi} L_{pd}\Rightarrow L_{pd}\approx\frac {a_{0}^2}{\epsilon^2} \frac {\omega_{0}^2}{\omega_{p}^2}c\tau\approx\frac{1}{4} a_{0} \frac {\omega_{0}^2}{\omega_{p}^2}c\tau\sim a_{0}L_{etch} $ which does depend on $a_{0}$. Using this scaling for $L_{acc}$ in equation (\ref{energygain}) provides the same scaling for the particle energy as obtained from similarity theory. However, even for this high intensity limit dephasing will still occur because the leading edge cannot move faster than the linear group velocity of the highest frequency component, since at the very front the amplitude is small \cite{Decker94}.

Last, it is useful to extrapolate the bubble regime for moderate $a_0$ to 10GeV and beyond.  If we keep $P/P_c$ fixed, then the laser power and inverse of the density scale with the desired electron energy and the pulse length will scale as the electron energy to the 1/2 power. If we scale up our sample simulation,  to obtain $15GeV$ we need a $2 PW$, $100fs$ laser and a plasma with density of $1.5 \times 10^{17} cm^{-3}$.  The only uncertainty is whether self-guiding is still possible. 
 
\begin{figure}[!tb]
\begin{center}
\includegraphics[width=3.4in]{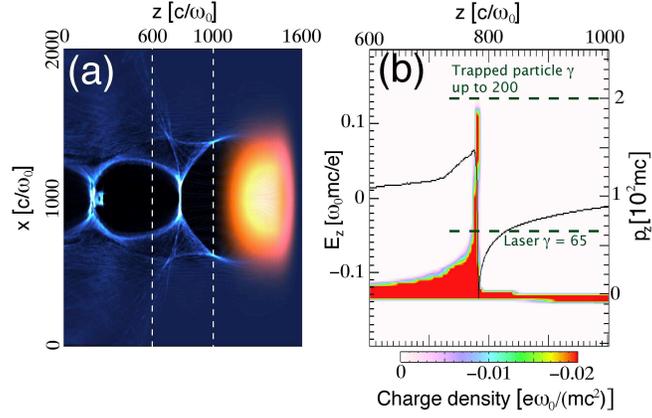}
\caption{Picture (a) shows the laser electric field with orange on top of the electron density with blue after $0.4mm$ of propagation. The laser front is chosen sharper than the back because it has been found through simulations that this leads to more stable propagation. The blowout region is well formed. Picture (b) shows that particles at the rear of the ion channel have already reached velocities higher than the laser velocity and therefore are trapped. The region in $z$ direction plotted in picture (b) corresponds to the $z$ region between the broken lines in picture (a).} \label{figure5}
\end{center}
\end{figure}

 To obtain condition for self-guiding  we resort to quasi-static theory where the index of refraction $\eta \simeq 1 - \frac{1}{2}\frac{\omega_p^2}{\omega_0^2}\frac{1}{1+\psi}$ and $\frac{\partial^2 \psi}{\partial\xi^2}- \frac{k_p^2}{2}\biggl[\frac{1+a^2}{(1+\psi)^2}-1\biggr] = 0$. As noted earlier  $1-\frac{1}{1+\psi} \simeq\psi$ must equal $4/(k_p w_0)^2$ in order for guiding to occur. For $|\psi| \ll 1$, we assume $\frac{\partial^2 \psi}{\partial\xi^2} \simeq\frac{k_p^2}{2}a^2$; therefore $\psi$ builds up to the necessary value in a distance $\Delta \xi$ which scales as $\Delta \xi \sim (k_{p}^{2}w_{0} a_{0})^{-1} \sim (k_{p} a_{0}^{3/2})^{-1}$. The length of the laser that is lost due to diffraction each Rayleigh length ($Z_{R}$) will also scale as $(k_{p} a_{0}^{3/2})^{-1}$ . Self guiding can be achieved if the length lost due to pump depletion in each Rayleigh length also scales the same way:
\begin{equation} \label{self_guiding}
\frac{\upsilon_{etch}}{c} \simeq \frac{3}{2}\frac{n_p}{n_c}  \sim \frac{\Delta \xi}{Z_{R}} \Rightarrow a_{0}\sim(n_{c}/n_{p})^{1/5}
\end{equation}
This scaling law indicates  that as the density is decreased to increase the electron energy we need to slowly increase the laser intensity to maintain self-guiding.   As a result the  $2PW$ laser pulse estimate needs to be slightly modified. Instead we  choose $P/P_{c}\simeq 28\Rightarrow a_{0}\simeq 6 $, and a density $n = 4\times 10^{17} cm ^{-3}$. For these parameters stable self guiding should occur and lead to acceleration of $1.8nC$ charge at approximately $9GeV$ after  $10 cm$ of laser propagation.  Carrying out a full-scale PIC simulation for the 10cm of propagation distance is beyond current capabilities. However, we have carried out a 3D PIC simulation for a short propagation distance to verify that electrons will still be self-injected. Results are shown in figure \ref{figure5}. 

If both the technology for making meter scale ($0.1m\sim1m$) low density plasma channels for guiding and for synchronized external injection are developed, then much higher energy gain can be achieved for the same laser power. For example, using $P/P_c\simeq 1$ (and the channel depth $\Delta{n}_c/n_p=0.5$), a $10GeV$ electron beam with  $\sim 0.6nC$ charge could be obtained with a $250TW$, $100fs$ laser and a $43cm$ long plasma channel with a minimum density $n = 1.2\times10^{17} cm^{-3}$. Extrapolating the parameters furrther, we predict that a $120GeV$ electron beam with $2.2nC$ charge generated by using a $3PW$, $350fs$ laser  and $18m$ long plasma channel with density $n = 1.0\times10^{16} cm^{-3}$.

The work is supported by DOE under grants DE-FC02-01ER41179, De-FG02-03ER54721, DE-FG03-92-ER4727 and  DE-FG03-NA0065,  by NSF under grant Phy-0321345, and by FCT (Portugal).   The simulations are performed on Dawson cluster, maintained locally by UCLA/ATS, and on the IBM SP @ NERSC under mp113 and gc2.  We acknowledge very useful conversations with Profs. T. Katsouleas and A. Pukhov.

\section{APPENDIX: PARAMETERS and SETUP}
The simulation for the $200TW$, $30fsec$ laser pulse in a fully ionized $n=1.5\times 10^{18}cm^{-3}$ plasma has a computational window of dimension $101.9
\times 127.3 \times 127.3 \mu m^{3}$ which moves at the speed of
light. The number of gridpoints is $4000\times256\times256 = 2.62
\times 10^{8}$.  The resolution in the laser propagation direction
$z$ is $k_{0}\Delta z = 0.2$.  We assume a preformed fully ionized
plasma with uniform density profile. The resolution in the
transverse direction is $k_{p}\Delta x = k_{p}\Delta y = 0.116$.
We use $2$ electrons per cell and a smooth neutralizing immobile
ion background (the total number of particles is roughly $500$
million).  A diffraction limited, circularly polarized pulse is
focused at the plasma entrance with a spot size $w_{0} = 19.5 \mu
m$ and the electric field has a symmetric temporal profile of
$10\tau^{3}-15\tau^{4} + 6\tau^{5}$ where $\tau =
\sqrt{2}(t-t_{0})/\tau_{FWHM}$.  The laser is Gaussian in the
transverse direction.  The total axial length of the plasma is
close to $0.75 cm$, or $300,000$ simulation time-steps.

The  simulation for the $2PW$, $100fsec$, $w_0 = 40 \mu m$ laser  in plasma density $n=4\times 10^{17}$ has  a computational window of dimension $203.8
\times 127.3 \times 127.3 \mu m^{3}$ which corresponds to $8000\times256\times256 = 5.24
\times 10^{8}$ gridpoints with identical resolution as in the aforementioned $200TW$ run. Roughly $1$ billion particles were used and ran $16000$ simulation time-steps.  The rise-time of the laser was about $2.5$ times shorter than the fall-time.

\end{document}